\documentstyle{mn}
\input{psfig}

\newif\ifAMStwofonts

\ifoldfss
  \ifCUPmtlplainloaded \else
    \NewTextAlphabet{textbfit} {cmbxti10} {}
    \NewTextAlphabet{textbfss} {cmssbx10} {}
    \NewMathAlphabet{mathbfit} {cmbxti10} {} 
    \NewMathAlphabet{mathbfss} {cmssbx10} {} 
  \fi
  \ifAMStwofonts
    \ifCUPmtlplainloaded \else
      \NewSymbolFont{upmath} {eurm10}
      \NewSymbolFont{AMSa} {msam10}
      \NewMathSymbol{\upi}     {0}{upmath}{19}
      \NewMathSymbol{\umu}     {0}{upmath}{16}
      \NewMathSymbol{\upartial}{0}{upmath}{40}
      \NewMathSymbol{\leqslant}{3}{AMSa}{36}
      \NewMathSymbol{\geqslant}{3}{AMSa}{3E}

      \let\leq=\leqslant \let\le=\leqslant
      \let\geq=\geqslant 
    \fi
  \fi
\fi 

\ifnfssone
  \newmathalphabet{\mathit}
  \addtoversion{normal}{\mathit}{cmr}{m}{it}
  \addtoversion{bold}{\mathit}{cmr}{bx}{it}
  \newmathalphabet{\mathbfit} 
  \addtoversion{normal}{\mathbfit}{cmr}{bx}{it}
  \addtoversion{bold}{\mathbfit}{cmr}{bx}{it}
  \newmathalphabet{\mathbfss} 
  \addtoversion{normal}{\mathbfss}{cmss}{bx}{n}
  \addtoversion{bold}{\mathbfss}{cmss}{bx}{n}
  \ifAMStwofonts
    \ifCUPmtlplainloaded \else
      \UseAMStwoboldmath
      \makeatletter
      \new@mathgroup\upmath@group
      \define@mathgroup\mv@normal\upmath@group{eur}{m}{n}
      \define@mathgroup\mv@bold\upmath@group{eur}{b}{n}
      \edef\UPM{\hexnumber\upmath@group}
      \new@mathgroup\amsa@group
      \define@mathgroup\mv@normal\amsa@group{msa}{m}{n}
      \define@mathgroup\mv@bold\amsa@group{msa}{m}{n}
      \edef\AMSa{\hexnumber\amsa@group}
      \makeatother
      \mathchardef\upi="0\UPM19
      \mathchardef\umu="0\UPM16
      \mathchardef\upartial="0\UPM40
      \mathchardef\leqslant="3\AMSa36
      \mathchardef\geqslant="3\AMSa3E

      \let\leq=\leqslant \let\le=\leqslant
      \let\geq=\geqslant 
    \fi
  \fi
\fi 

\ifnfsstwo
  \DeclareMathAlphabet{\mathbfit}{OT1}{cmr}{bx}{it}
  \SetMathAlphabet\mathbfit{bold}{OT1}{cmr}{bx}{it}
  \DeclareMathAlphabet{\mathbfss}{OT1}{cmss}{bx}{n}
  \SetMathAlphabet\mathbfss{bold}{OT1}{cmss}{bx}{n}
  \ifAMStwofonts
    \ifCUPmtlplainloaded \else
      \DeclareSymbolFont{UPM}{U}{eur}{m}{n}
      \SetSymbolFont{UPM}{bold}{U}{eur}{b}{n}
      \DeclareSymbolFont{AMSa}{U}{msa}{m}{n}
      \DeclareMathSymbol{\upi}{0}{UPM}{"19}
      \DeclareMathSymbol{\umu}{0}{UPM}{"16}
      \DeclareMathSymbol{\upartial}{0}{UPM}{"40}
      \DeclareMathSymbol{\leqslant}{3}{AMSa}{"36}
      \DeclareMathSymbol{\geqslant}{3}{AMSa}{"3E}

      \let\leq=\leqslant \let\le=\leqslant
      \let\geq=\geqslant 
    \fi
  \fi
\fi 

\ifCUPmtlplainloaded \else
  \ifAMStwofonts \else 
    \def\upi{\pi}
    \def\umu{\mu}
    \def\upartial{\partial}
  \fi
\fi

\def\LaTeX{L\kern-.36em\raise.3ex\hbox{a}\kern-.15em
    T\kern-.1667em\lower.7ex\hbox{E}\kern-.125emX}

\title[Binary second sequences in cluster CMDs]{The Binary Second
Sequence in Cluster Colour--Magnitude Diagrams}

\author[J. Hurley and C. A. Tout]
  {Jarrod Hurley and Christopher A. Tout\\
Institute of Astronomy, The Observatories, Madingley Road,
Cambridge CB3 0HA}

\date{Accepted ... . Received ...; in original form 
...}

\pagerange{\pageref{firstpage}--\pageref{lastpage}}
\pubyear{1998}

\begin{document}

\label{firstpage}

\maketitle

\begin{abstract}
We show how the second sequence seen lying above the main sequence in
cluster colour magnitude diagrams results from binaries with a large
range of mass ratios and not just from those with equal masses.  We
conclude that the presence of a densely populated second sequence, with
only sparse filling in between it and the single star main sequence, does
not necessarily imply that binary mass ratios are close to unity.
\end{abstract}

\begin{keywords}
 stars: clusters of -- stars: binary
\end{keywords}

\section{Introduction}

With the recent surge of high quality colour--magnitude diagrams
(hereinafter CMDs) of
clusters obtained both from space (Rubenstein \& Bailyn 1997, Richer {\it{et al}}. 
1997 and Elson {\it{et al}}. 1998) and ground based 
(Ferraro {\it{et al}}. 1997)
observations it is timely to reconsider the effect of binary stars on
the observed stellar main sequence.  
It is well known that an
unresolved binary system comprising two identical stars has the same
colour but twice the luminosity of an equivalent single star and that 
such a system, comprising two equal-mass main-sequence stars,
will appear in the cluster CMD displaced vertically by 0.753
magnitudes irrespective of the wavelength bands used (Haffner \& Heckmann 1937).
Because of
this, the clear second sequences displaced by this amount above the
main-sequence visible in many cluster CMDs are taken to indicate a
population of equal-mass binaries (Bergbusch {\it{et al}}. 1991, Kaluzny \& Rucinski 1995 and Santiago {\it{et al}}. 1996) .  
The fact that the region between
the single-star main sequence and this second sequence is not
very densely populated is then often taken to mean that the
mass-ratio distribution in the binary systems is biased towards equal
masses.  However it turns out that this need not be the case.  A system
with two unequal main-sequence components has a combined colour that
is redder than the colour of the brighter component as well as a
luminosity greater than the single star but less than the
corresponding equal-mass binary (Bolte 1991 and Romani \& Weinberg 1991).  
Such a system is displaced both upwards and to the
right relative to the main-sequence position of the brighter component.
Thus if we consider the position of a system with a mass ratio
slightly different from unity, relative to the equal-mass system it
moves to fainter magnitude but also to redder colour.  Now, because
the intrinsic slope of the main sequence is such that less massive
stars are fainter and redder, it is possible for systems with unequal
components to follow the second sequence downwards and to the right in
the CMD.  We show (section~2) that, for certain mass ranges and choice of colours,
systems with quite extreme mass ratios still lie close to the second
sequence, $0.753\,$mag above the actual main sequence, appropriate to
their combined colour.  We then show (section~3) how an unbiased mass
ratio distribution can lead to the clearly separated main and second
sequences observed in cluster CMDs.

\section{Combining the Colours and Magnitudes}

For a given colour index $X$ a main-sequence star of bolometric
luminosity $L$ has an absolute magnitude
\begin{equation}
M_X = M_{X,\odot} - 2.5\log_{10}{L\over L_\odot} + \beta_X(T_{\rm
eff}),
\end{equation}
where $\beta_X$ is the bolometric correction appropriate to a
main-sequence star with effective temperature $T_{\rm eff}$ and
$M_{X,\odot}$ and $L_\odot$ are the absolute magnitude and bolometric
luminosity of the Sun.  Given a second colour $Y$ a corresponding
absolute magnitude $M_Y$ can be defined in a similar way and a colour
by
\begin{equation}
(X - Y) = M_X - M_Y.
\end{equation}
In the case of an unresolved binary the two components contribute
differently to $M_X$ and $M_Y$ because of their different effective
temperatures and consequently different bolometric corrections.  Using
the suffices 1 and~2 for the components and~3 for the system we can
write the combined magnitudes as
\begin{equation}
M_{X,3} = M_{X,\odot} - 2.5\log_{10}{L_{X,1} + L_{X,2}\over L_\odot},
\end{equation}
where
\begin{equation}
L_{X,i} = L_i 10^{-0.4 \beta (T_{{\rm eff},i})} .
\end{equation}
The combined colour is then
\begin{equation}
(X - Y)_3 = M_{X,3} - M_{Y,3}.
\end{equation}
For a binary system of total mass $M = M_1 + M_2$ and mass ratio $q =
M_2/M_1$, where we choose the primary mass $M_1 > M_2$ (corresponding
to $L_1 > L_2$ on the main sequence) and hence $0\le q\le 1$, we can
calculate both $L_i$ and $T_{{\rm eff},i}$ by means of the zero-age
main-sequence formulae constructed by Tout et al. (1996).  We can then
use the synthetic stellar spectra computed by Kurucz (1992) 
to find the appropriate bolometric
corrections $\beta_{X,i}$ needed to generate a theoretical CMD.  
Figure 1 shows such a CMD containing the zero-age single star and equal-mass 
 binary sequences for solar metallicity. 
Also shown for a range of primary masses are the locus of binary points with $q$ 
ranging from $1.0 \rightarrow 0.0$ in increments of 0.1. 
We see that, as $M_2$ is decreased for a particular $M_1$, the binary becomes 
fainter but also redder for $q$ greater than some $q_{\rm{crit}}$.
Thus the locus of binary points follows the equal-mass sequence for mass ratios  
in the range $q_{\rm{crit}} < q \leq 1$. 
For $M_1 \geq 3\, M_{\odot}$ we find $q_{\rm{crit}} \simeq 0.5$ while for 
$M_1 \leq 2\, M_{\odot}$ we find $q_{\rm{crit}} \simeq 0.9$.
These ideas were originally formulated by Haffner \& Heckmann (1937) and later developed by Bettis (1975) and Dabrowski \& Beardsley (1977) but neither of the latter two make any connection between magnitude difference and mass ratio.
Our results, based on up to date stellar models and bolometric corrections, represent a thorough reworking.

\begin{figure}
\psfig{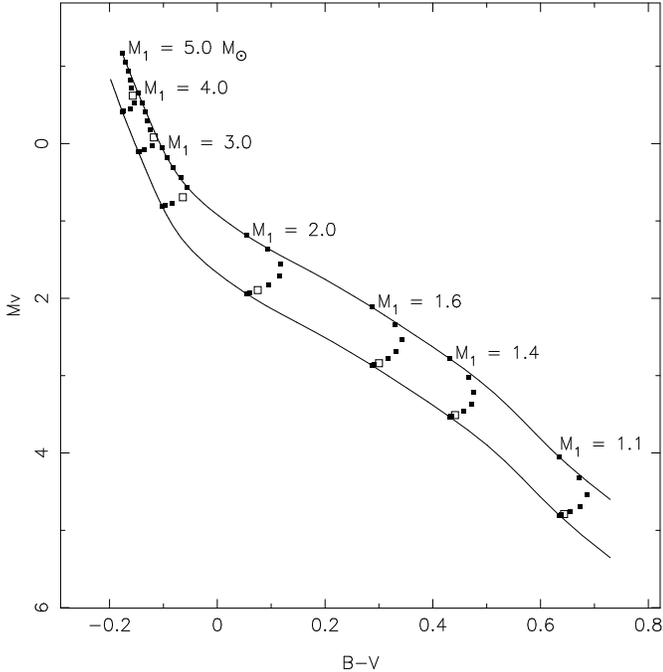}
\caption{Theoretical zero-age single star and equal-mass binary main
sequences for a metallicity of $Z = 0.02$  covering a range of $M_V$.
Also plotted for a range of primary masses, $M_1$, are binary points with
 $M_2 = q M_1$ where q ranges
from $1.0 \rightarrow 0.0$ in increments of 0.1. The point at $q = 0.5$ is
an open square.}
\end{figure}

\section{Cluster Colour Magnitude Diagrams}

To understand how these effects manifest themselves in observed CMDs
we consider a distribution of 5000 stars of which 4000 are in unresolved binary
systems.  We select the masses of the single stars and of the binary
primaries according to the initial mass function derived by Kroupa,
Tout \& Gilmore (1993) by means of the generating function
\begin{equation}
{{M_1} \over {M_{\odot}}} = 0.08 + {{0.19 X^{1.55} + 0.05 X^{0.6}} \over 
{\left( 1.0 - X \right)^{0.58}}},
\end{equation}
where we choose X uniformly distributed to give
masses in the range $1 < M_1 / M_{\odot} < 6$.  
We then choose the mass ratio
in the binary systems so that $q$ is uniformly distributed between $0$
and $1$ to find the secondary masses
\begin{equation}
M_2 = qM_1.
\end{equation}
For each single star and binary system we calculate the colours and
magnitudes as described in the previous section and plot them in the
CMD shown in Figure~2.
We see that the densest population is close to the single star main sequence 
even though most stars are binary. 
This is because faint secondaries make no contribution. 
With this particular mass ratio distribution we see a clear second sequence 
for $ \left( B - V \right) < 0.1$ corresponding to $M_1 > 2\,M_{\odot}$.
This is much clearer in the expanded region shown in Figure~2b than in the 
full CMD in Figure~2a and we note that this is due to the size 
of the symbols used to plot the stars relative to the space between them and care 
should be taken when interpretating observations by eye. 
This point CMD is in the form in which observations have been presented so 
far but we note that a density contour plot would be less open to false 
interpretation (see Figure~3). 
Between $0.1 < \left( B - V \right) < 0.3$ no clear second sequence is 
visible but for $\left( B - V \right) > 0.3$ it is again apparent in 
Figures~2c and~3c.

\section{Discussion}

\begin{figure*}
\psfig{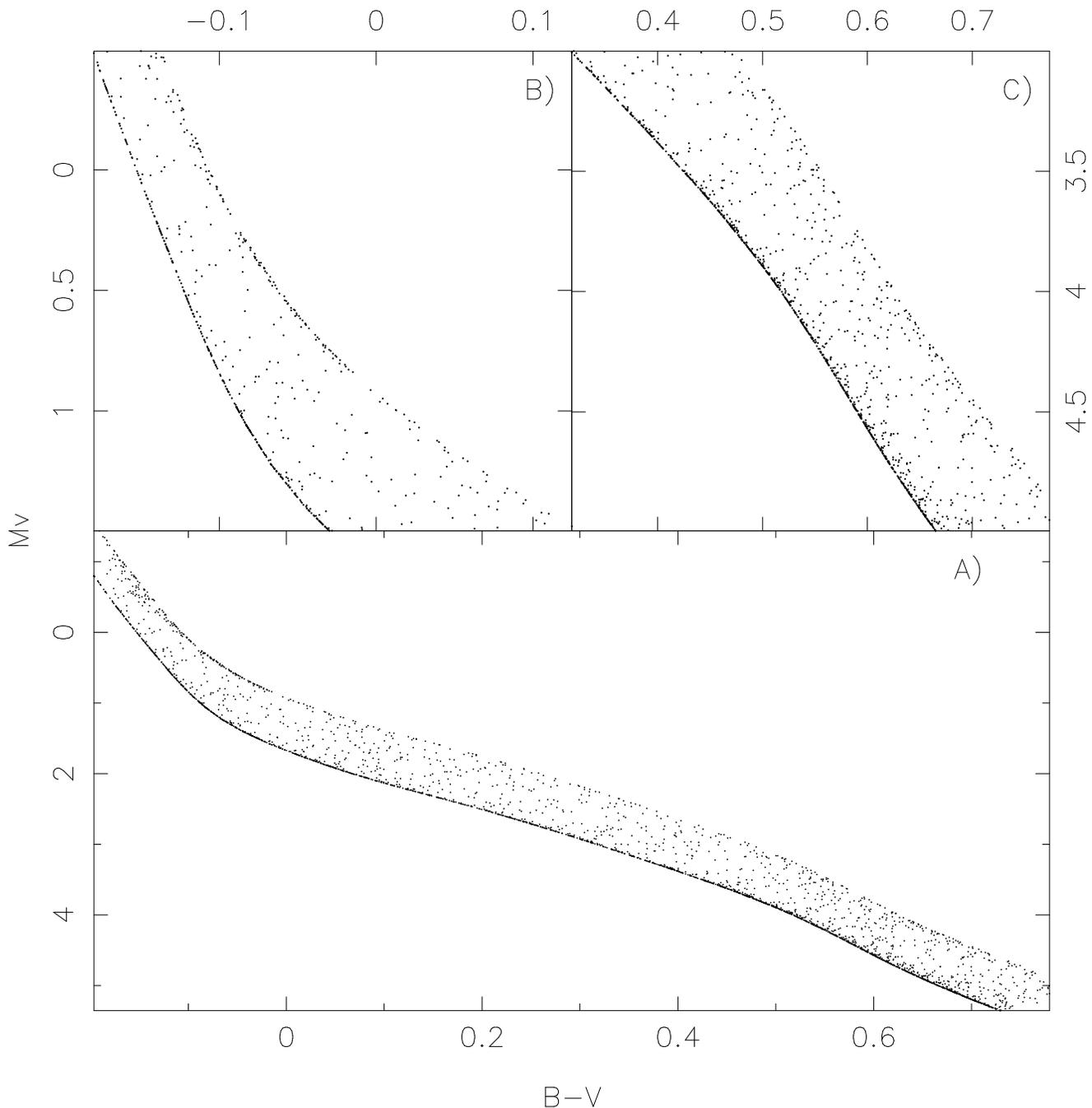}
\caption{ A) CMD for a randomly generated zero-age population of binary 
and single stars. B) and C) show two enlarged portions of the lower panel 
which correspond to approximate mass ranges $2 < M_1 < 4$ and 
$1 < M_1 < 1.4$ respectively, both of which show well-defined second 
sequences.}
\end{figure*}

We have shown that not only equal mass main-sequence binary systems lie
on the second sequence $0.753\,$mag above the single star sequence but
also many with unequal masses.  In particular for $B$ and $V$ colours
and primary masses $M_1 > 2$ even systems with quite extreme
mass ratios, as low as $q = 0.5$, lie on the second sequence.  We have also
shown how a cluster with uncorrelated binary masses can show a
distinct second sequence in the CMD.  However we must comment on what
we mean by uncorrelated.  For Figures~2 and~3 we have selected each primary
mass $M_1$ from the IMF and then chosen $q < 1$ uniformly distributed.
Note that choosing $q_{\max} > q > 1$ uniformly would lead to quite a
different distribution and it can be argued that we really should be
choosing $0 > \log q > \log q_{\min}$ uniformly.  Alternatively
Eggleton, Fitchett \& Tout (1989) defined uncorrelated to mean the two masses
chosen independently from the same initial mass function.  For a
nonuniform initial mass function such a choice does not lead to a
uniform distribution for $q$ overall nor for any particular primary
mass.  Indeed, because any star is most likely to have a low-mass
companion, systems with massive primaries will tend to have extreme
mass ratios while those with low-mass primaries will tend to have
mass ratios closer to unity.  In this sense our uniform distribution
in $0 < q < 1$ actually corresponds to a correlation in the masses for
primaries with $M_1 > 2\,M_\odot$.  A CMD with masses chosen independently
from the initial mass function given above does not have a clear
second sequence around $3\,M_\odot$ where extreme mass ratios are more
likely but does at low masses where
systems tend to have equal mass components.  To distinguish between
the possible mass ratio distributions we note that we can employ
statistical methods such as those described by Kroupa and Tout (1992)
to investigate the binary content of Praesepe.  In that work the
colours and magnitudes were implicitly combined according to the
prescription of section~2.

\section*{ACKNOWLEDGMENTS}

CAT is very grateful to PPARC for support from an advanced
fellowship. 
JH thanks Trinity College and the Cambridge Commonwealth 
Trust for their kind support.
We thank Sverre Aarseth, Steinn Sigurdsson, Melvyn Davies
and Rebecca Elson for the stimulating questions that have led to 
this contribution.

\begin{figure*}
\psfig{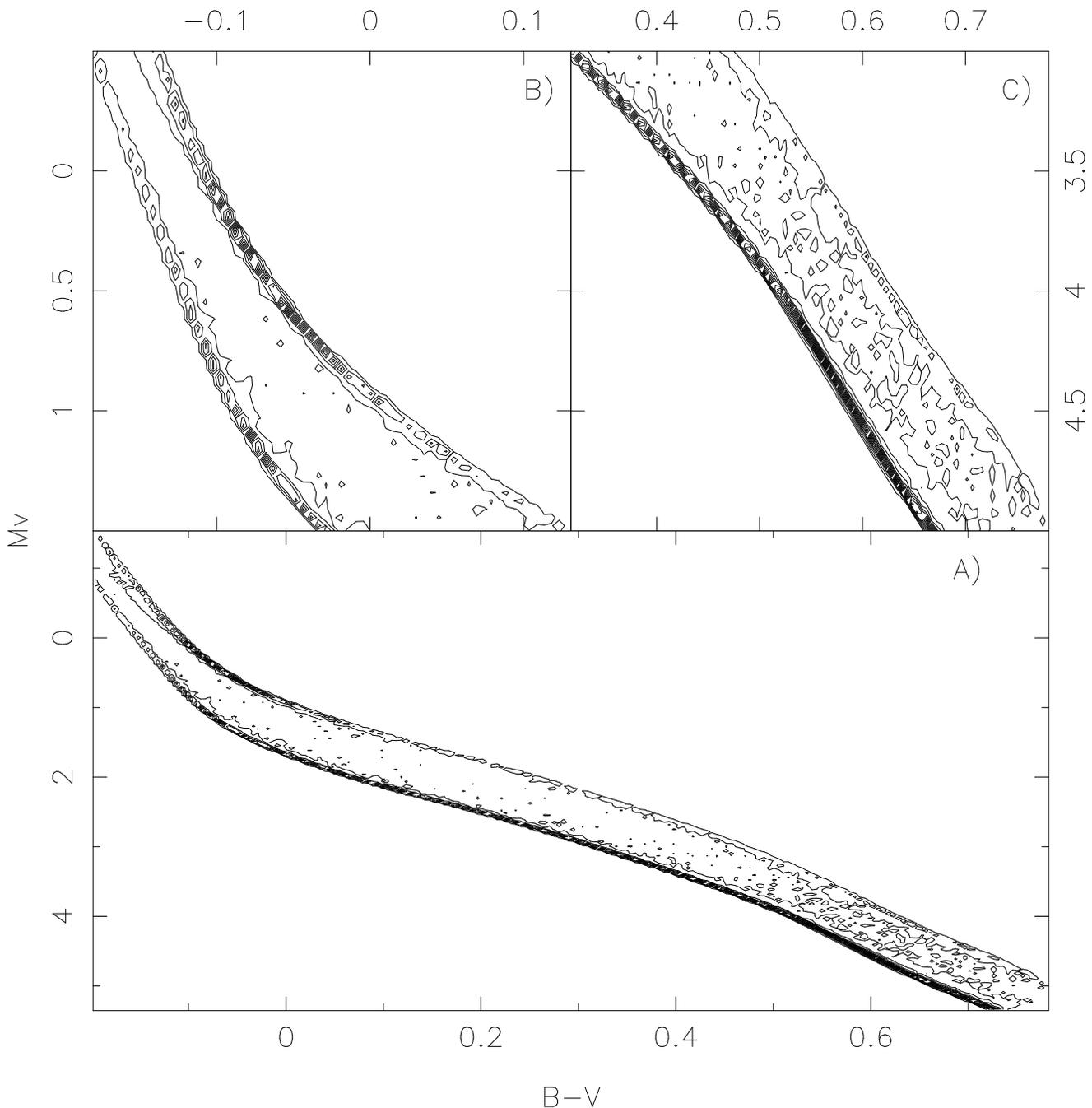}
\caption{ Number density plot corresponding to the same regions
 as the CMD of figure~2. $1\,000\,000$ randomly generated binaries are 
considered with no single stars. The grid is $200 \times 200$ with 
10 contour levels evenly spaced between 0.05 and 0.95 after 
normalisation of the data to the maximum density.}
\end{figure*}

\label{lastpage}

\end{document}